\documentclass{PoS}
\usepackage{amsmath,bm}
\pdfoutput=1	

\definecolor{plotpink}{RGB}{205,0,180}
\definecolor{plotcyan}{RGB}{0,215,215}
\definecolor{plotblue}{RGB}{0,0,235}
\definecolor{plotorange}{RGB}{245,140,0}
\definecolor{plotgreen}{RGB}{30,130,0}
\definecolor{plotred}{RGB}{240,0,0}
\definecolor{darkgreen}{RGB}{0,170,0}

\newcommand{\cmsqs}{\mathrm{cm}^3\mathrm{s}^{-1}}
\newcommand{\mDM}{m_\mathrm{DM}}
\newcommand{\GeV}{\mathrm{GeV}}
\renewcommand{\vec}[1]{\boldsymbol{#1}}
\newcommand{\diff}{\mathrm{d}}

\newcommand{\sv}{\langle\sigma v \rangle}
\newcommand{\J}{J_\mathrm{GC}}

\title{
A combined dark matter study of AMS-02 antiprotons and Fermi-LAT gamma rays 
}

\ShortTitle{A combined DM study of AMS-02 antiprotons and Fermi-LAT gamma rays}
\author{}
\author{Alessandro Cuoco, \speaker{Jan Heisig}, 
      Michael Korsmeier and Michael Kr\"amer
      \\
        Institute for Theoretical Particle Physics and Cosmology,
RWTH Aachen University, 52056 Aachen, Germany\\
        E-mail: \email{cuoco@physik.rwth-aachen.de},
\email{heisig@physik.rwth-aachen.de}, 
\email{korsmeier@physik.rwth-aachen.de}, 
\email{mkraemer@physik.rwth-aachen.de}}

\abstract{
Observations of cosmic rays are a sensitive probe of dark matter annihilation in our Galaxy.
In this article we present an analysis of the AMS-02 antiproton data, reducing cosmic-ray 
propagation uncertainties by fitting at the same time dark matter and propagation parameters.
The result exhibits a possible hint for dark matter pointing to an annihilation cross section 
close to the thermal value. We investigate the compatibility of this signal with a dark matter 
interpretation of the Galactic center excess seen in the Fermi-LAT gamma-ray data and 
discuss implications for dark matter models.
}

\FullConference{EPS-HEP 2017, European Physical Society conference on High Energy Physics\\
		5-12 July 2017\\
		Venice, Italy}

\begin{document}

\section{Introduction}\label{sec:intro}

The nature of dark matter (DM) in the Universe constitutes one of the most pressing questions in contemporary
fundamental physics. It has motivated an impressive joint effort in the field of particle- and astrophysics.
Among the various strategies, indirect detection searches are of particular importance as they
probe the self-annihilating nature of DM, potentially establishing a thermal freeze-out scenario as
an explanation of the measured DM relic density. With the recently published data of the AMS-02
experiment on-board the international space station the cosmic-ray (CR) precision era has begun, 
allowing for a significant improvement in the sensitivity of DM searches. However, DM constraints
are largely affected by uncertainties in the description of CR propagation in the Galaxy. 
This results in an order-of-magnitude uncertainty in the DM interpretation of CR fluxes 
when derived for benchmark propagation models, like the MIN/MED/MAX 
scenarios~\cite{Donato:2003xg}.

Here, we summarize our recent work~\cite{Cuoco:2016eej,Cuoco:2017rxb} where we exploit the 
precise AMS-02 measurements of the fluxes of protons, helium and antiprotons. By preforming 
a joint fit of the propagation parameters and DM signal we minimize the CR propagation 
uncertainties. This treatment explores possible correlations and degeneracies between 
CRs and DM, providing more robust and reliable DM constraints. While being able to put 
strong constraints on the annihilation cross section at large masses, we find a statistically 
significant (around $4.5\sigma$ local) indication of a signal from DM annihilation for lower 
masses $\mDM\lesssim100\,\GeV$ pointing to an annihilation cross section close to the 
thermal one, $\langle\sigma v \rangle \simeq 3\times 10^{-26}\,\cmsqs$. We analyse 
this indication for a variety of annihilation channels. Intriguingly, this possible signal from 
DM is largely compatible with the Galactic center excess (GCE) seen in the Fermi-LAT 
gamma-ray data.

\section{Cosmic-ray analysis}

The propagation of charged CRs can be described by a diffusion 
equation~\cite{StrongMoskalenko_CR_rewview_2007} for the 
particle density $\psi_i$ of species $i$ per volume and absolute 
value of momentum $p$
\begin{equation}
\begin{split}
  \label{eqn::PropagationEquation}
  \frac{\partial \psi_i (\vec{x}, p, t)}{\partial t} = \;&
    q_i(\vec{x}, p) +  
    \vec{\nabla} \cdot \left(  D_{xx} \vec{\nabla} \psi_i - \vec{V} \psi_i \right)
     +  \frac{\partial}{\partial p} p^2 D_{pp} \frac{\partial}{\partial p} \frac{1}{p^2} \psi_i   \\
   &- \frac{\partial}{\partial p} \left( \frac{\diff p}{\diff t} \psi_i  - \frac{p}{3} (\vec{\nabla \cdot V}) \psi_i \right) 
    - \frac{1}{\tau_{f,i}} \psi_i - \frac{1}{\tau_{r,i}} \psi_i\,,
    \end{split}
\end{equation} 
where $q_i(\bm{x}, p)$ is the source term of primary CRs and the terms 
proportional to $D_{xx},  \bm{V}$, and $D_{pp}$ correspond to CR diffusion, 
convection and reacceleration, respectively. The second line of 
eq.~\eqref{eqn::PropagationEquation} includes the momentum gain or 
loss rate~$\propto {\diff p}/{\diff t}$, adiabatic energy losses 
$\propto \bm{\nabla \cdot V}$, and the loss by fragmentation and radioactive 
decay $\propto 1/\tau_{f,i}$ and $1/\tau_{r,i}$, respectively. 
We solve eq.~\eqref{eqn::PropagationEquation} numerically using 
\textsc{Galprop}~\cite{Strong:1998fr} approximating the Galaxy by 
a cylindrical volume assuming a steady state regime. 
The source term for primary astrophysical CRs is modeled by a 
double broken power law with smooth transitions. We allow for individual 
spectral indices for proton and helium. Diffusion is modeled by a power law 
in rigidity and convective winds are assumed to be orthogonal to the Galactic 
plane. The respective parametrizations are detailed in~\cite{Korsmeier:2016kha} 
and introduce a total of 11 background propagation parameters. 

The source term for primary antiprotons from DM annihilation can be written as:
\begin{eqnarray}
  \label{eqn::DM_source_term}
  q_{\bar{p}}^{(\mathrm{DM})}(\bm{x}, E_\mathrm{kin}) = 
  \frac{1}{2} \left( \frac{\rho(\bm{x})}{m_\mathrm{DM}}\right)^2  
  \sum_f \left\langle \sigma v \right\rangle_f \frac{\diff N^f_{\bar{p}}}{\diff E_\mathrm{kin}} ,
\end{eqnarray}
where $m_\mathrm{DM}$ denotes the DM mass, $\rho(\bm{x})$ the DM density 
profile, $\left\langle \sigma v \right\rangle_f$ the thermally averaged annihilation 
cross section and $\diff N^f_{\bar{p}}/\diff E_\mathrm{kin}$ the corresponding 
antiproton energy spectrum per DM annihilation, see~\cite{Cuoco:2016eej,Cuoco:2017rxb} 
for further details. The factor $1/2$ in eq.~\eqref{eqn::DM_source_term}
corresponds to Majorana fermion DM, the corresponding factor for a Dirac fermion is $1/4$.
For a given annihilation channel $f$, ${\rm DM}\!+\!{\rm DM} \to f\!+\!\bar{f}$, 
the source term introduces two additional parameters, $m_\mathrm{DM}$ and
$\left\langle \sigma v \right\rangle_f$.
The resulting 13 parameters are determined in a global fit of the AMS-02 
proton and helium fluxes~\cite{Aguilar_AMS_Proton_2015, Aguilar_AMS_Helium_2015}
and the AMS-02 antiproton-to-proton ratio \cite{Aguilar:2016kjl}, supplemented by 
proton and helium data from CREAM~\cite{Yoon_CREAM_CR_ProtonHelium_2011} and 
VOYAGER~\cite{Stone_VOYAGER_CR_LIS_FLUX_2013}. 
We use \textsc{MultiNest} \cite{Feroz_MultiNest_2008} for an efficient sampling. 
The VOYAGER data, directly measuring the local interstellar space flux, is used to 
constrain the solar modulation potential which we describe in the framework of 
the force-field approximation.

\section{Cosmic-ray results}

\begin{figure}[ht]
\centering
\setlength{\unitlength}{1\textwidth}
\begin{picture}(1,0.38)
  \put(-0.005,-0.015){\includegraphics[width=1.01\textwidth]{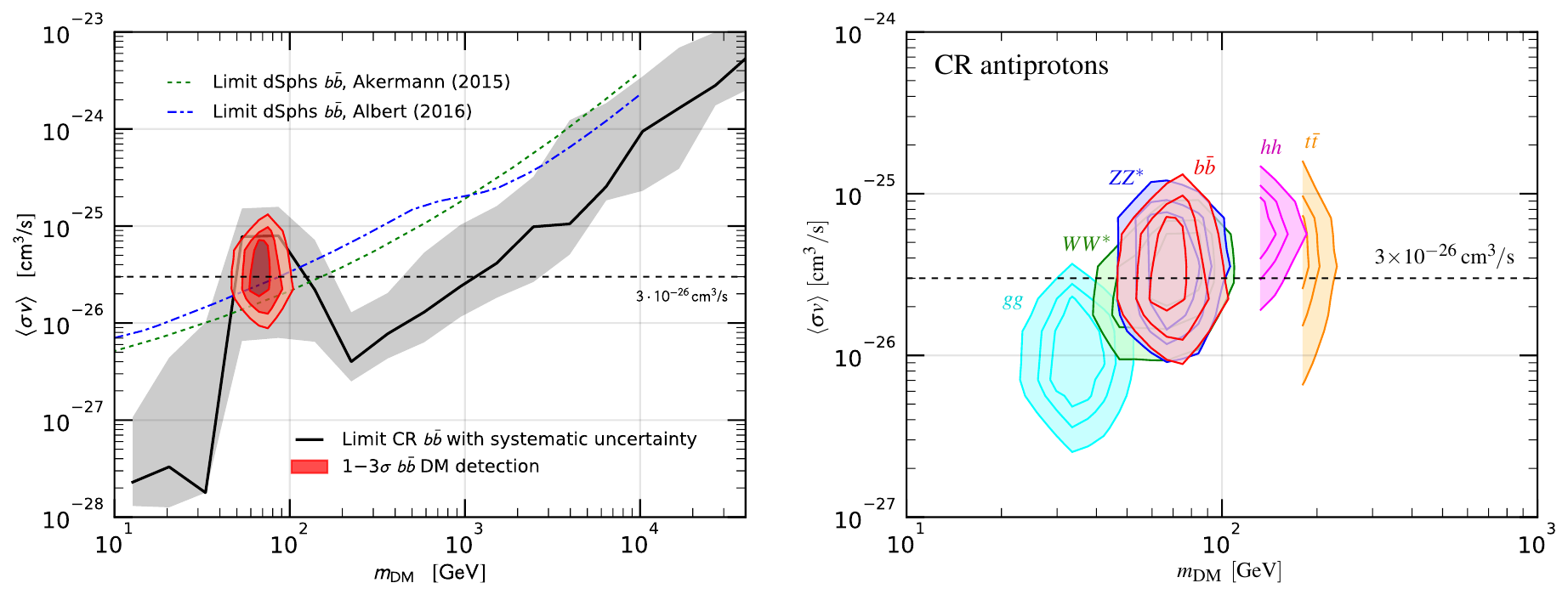}}
\end{picture}
\caption{Left:~Best fit region (1--3$\sigma$ contours) for a DM component of the 
antiproton flux, and limits on the DM annihilation cross section into $b\bar{b}$ 
final states; from~\cite{Cuoco:2016eej}. The grey shaded band denotes the 
systematic uncertainty. For comparison we show limits on the annihilation 
cross section obtained from gamma-ray observations of dwarf 
galaxies~\cite{Ackermann:2015zua,Fermi-LAT:2016uux}, and the thermal 
value of the annihilation cross section, 
$\left\langle \sigma v \right\rangle \simeq 3 \times 10^{-26}$~cm$^3$s$^{-1}$. 
Right:~Comparison of the best fit regions (1--3$\sigma$ contours) for the 
annihilation channels $gg$~(cyan), $WW^*$~(green), $b\bar b$~(red), 
$ZZ^*$~(blue), $hh$~(pink) and $t\bar t$~(orange); from~\cite{Cuoco:2017rxb}.
}
\label{fig:CRfit}
\end{figure}

We are able to put strong constraints on the annihilation cross section at large 
DM masses, see figure~\ref{fig:CRfit} (left) for the $b\bar b$ annihilation 
channel.\footnote{For CR antiproton constraints on a variety of other annihilation channels
see~\cite{Cuoco:2017iax}.} However, in the mass range around and below 
100\,GeV the global fit of the CR antiproton data improves significantly when 
adding a contribution of primary antiprotons from DM annihilation. This is due 
to a sharp spectral feature in the antiproton flux at a rigidity of about 20\,GV, which is
well fitted by the spectrum from DM annihilation due to the kinematic cut-off set by 
the DM mass. The feature cannot be described well by the secondary antiprotons 
produced by the interactions of primary protons and helium nuclei on the interstellar 
medium, since they only provide a smooth spectrum. 

Figure~\ref{fig:CRfit} (right) shows the best fit regions in the DM mass and 
annihilation cross section for the six considered annihilation channels. 
The regions are obtained by minimizing the $\chi^2$ with respect to the 
remaining 11 parameters in the fit. Hence, they include the uncertainties 
in the CR source spectra and CR propagation.
All channels provide an improvement compared to a fit without DM\@. 
For the latter we obtain a $\chi^2$/(number of degrees of freedom) of 71/165. 
This is reduced to 50/163\,($gg$),  50/163\,($WW^*$), 46/163\,($ZZ^*$)
46/163\,($b \bar b$), 48/163\,($hh$) and 59/163\,($t \bar t$), respectively, 
when adding a corresponding contribution from DM annihilation. 
Formally, $\Delta \chi^2 \simeq 25$ (obtained for $b \bar b$ and $ZZ^*$) 
corresponds to a statistical significance of $4.5\sigma$.\footnote{%
Note that this estimate does not account for possible systematic errors.} 
Figure~\ref{fig:CRfit} (right) also shows that different annihilation channels imply 
different preferred DM masses, ranging from $\mDM \simeq 35$\,GeV for gluons 
to $\mDM$ near the Higgs or top mass for annihilation into $hh$ or $t\bar t$, 
respectively. Note that for the channels $WW^*$ and $ZZ^*$ the best fit masses 
lie below the corresponding on-shell production thresholds. We therefore take 
into account the respective 3-body final state processes via one off-shell vector 
boson, denoted by the asterisk. For all channels the fit points to an annihilation 
cross section close to the value required for a thermal relic, 
$\left\langle \sigma v \right\rangle \simeq 3 \times 10^{-26}\,\cmsqs$. 

A deeper scrutiny of this potential signal would require a more accurate 
determination of the antiproton production cross section, an accurate 
modeling of solar modulation at low rigidities of less than about~5\,GV
as well as a better knowledge of correlations in the uncertainties of the 
AMS-02 data.

\section{Joint fit with Fermi-LAT gamma-ray data}

Annihilation of DM would also result in a continuum flux of gamma rays, 
predominantly from the decay of pions produced in 
the fragmentation of primary annihilation products. 
Accordingly, in this section we study whether the indication for a 
DM annihilation signal in CRs found in the last section is compatible 
with recent Fermi-LAT gamma-ray data. In particular, we consider data 
from dwarf spheroidal galaxies~\cite{Fermi-LAT:2016uux} and the 
so-called Galactic center excess (GCE) recently confirmed
by the Fermi-LAT collaboration~\cite{TheFermi-LAT:2015kwa}.

We consider a total of eleven dwarfs: the seven brightest confirmed dwarfs 
analyzed in~\cite{Fermi-LAT:2016uux} as well as Willman~1, Tucana~III, 
Tucana~IV and Indus~II. We use the public likelihood provided by 
Fermi-LAT~\cite{Fermi-LAT:2016uux} and marginalize over the $J$-factors 
according to their uncertainties. For the GCE we use the gamma-ray energy 
spectrum and error covariance matrix obtained in~\cite{Calore:2014xka}.
We then perform a joint global fit using the gamma-ray and CR likelihoods 
(marginalized over CR propagation parameters).
The gamma-ray fit follows the methodology described in~\cite{Cuoco:2016jqt}. 
Besides the two model parameters, $\langle \sigma v \rangle$ and $m_\text{DM}$, 
the fit introduces two nuisance parameters -- the $J$-factor for the Galactic center 
and the local DM density affecting the DM antiproton flux, which are varied around
$\log (\J/(\text{GeV}^{2}\text{cm}^{-5}))=53.54\pm 0.43$~\cite{Cuoco:2016jqt}
and $\rho_\odot=0.43\pm0.15$ GeV/cm$^3$~\cite{Salucci:2010qr}, respectively.
These two parameters take into account the uncertainty in the DM density profile
and can approximately be considered to be independent~\cite{Cuoco:2017rxb}.

\begin{figure}[t]
\centering
\setlength{\unitlength}{1\textwidth}
\begin{picture}(1,0.375)
  \put(-0.005,-0.015){\includegraphics[width=1.01\textwidth]{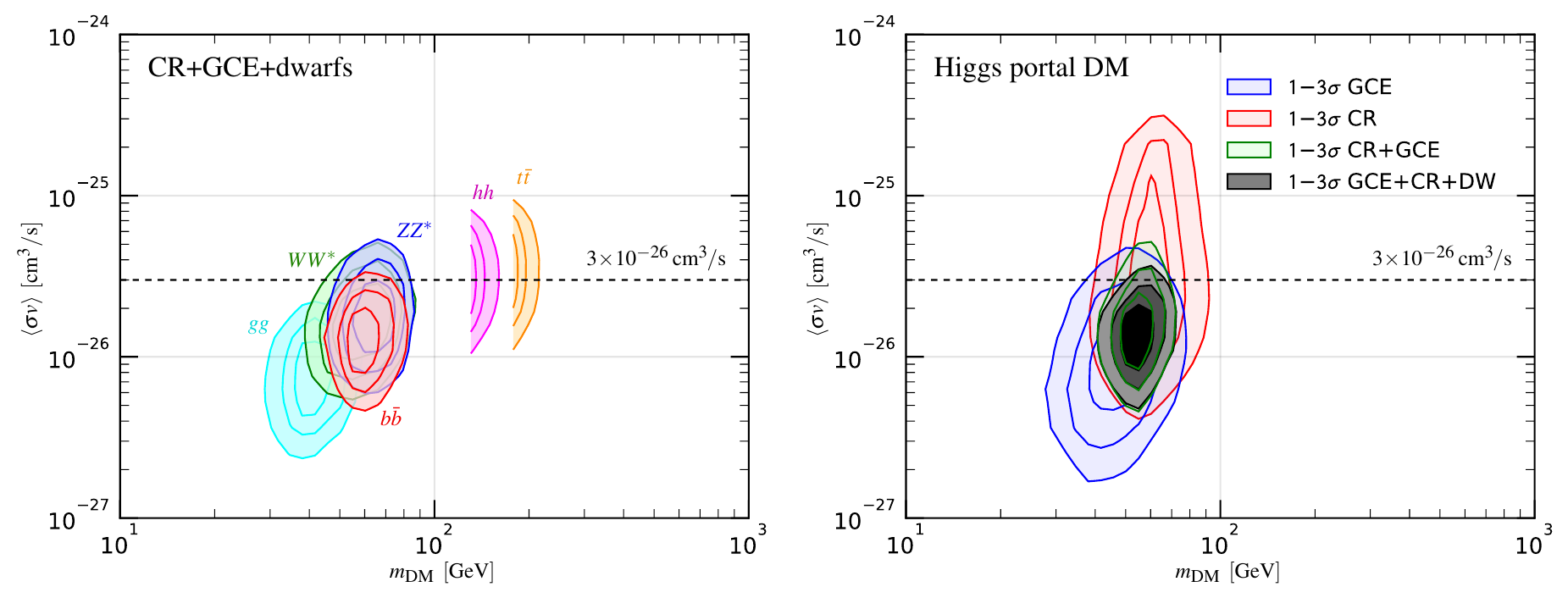}}
\end{picture}
\caption{Left: Same as figure~\protect\ref{fig:CRfit} (right) but taking into account the likelihood of Fermi-LAT dwarfs and the GCE in addition. Right: Individual and joint fits within the Higgs portal DM model; from~\cite{Cuoco:2017rxb}.}
\label{fig:genfit}
\end{figure}

Figure~\ref{fig:genfit} (left) shows the preferred range of DM masses and 
annihilation cross sections, where we have marginalized over $\J$ and 
$\rho_\odot$. For most annihilation channels, we observe very good 
agreement between the best fit regions of the CR signal and the DM 
interpretation of the GCE while both are in accordance with limits from 
dwarf spheroidal galaxies.
However, annihilation into gluons or top quarks is disfavored as an 
explanation of both the CR antiproton flux and the GCE. While both 
measurements individually prefer different regions of the DM mass for 
annihilation into gluons, annihilation into  $t\bar t$ does not provide a 
particularly good fit to both the GCE and the antiprotons.

\section{Implications for dark matter models}

Realistic DM models often feature an admixture of various annihilation channels. 
It is thus not clear \emph{a priori} if a certain combination of annihilation channels, 
considered individually in the last sections, can provide a good fit of the observations. 
Here we consider the Higgs portal model where a DM field $X$ is directly coupled to 
the Higgs bilinear through the operator $X^2H^\dagger H$. 
Independent of the nature of $X$, below the Higgs-pair production threshold,
the relative weights of the annihilation channels are solely determined by the DM mass.
Accordingly, we can provide a joint fit of the CR and gamma-ray data within the Higgs portal 
model in terms of the two parameters $\mDM$ and $\sv$. The result is shown in
figure~\ref{fig:genfit} (right), where we have, again, marginalized over $\J$ and $\rho_\odot$. 
The preferred region of DM masses is around $m_{\rm DM} \simeq 60$\,GeV, 
where the model predicts annihilation into $b\bar b$, $WW^*\!$ and $gg$ 
with a weight of approximately 70\%, 20\%, and 10\%, respectively. 
We find a $\chi_\mathrm{CR}^2$/(number of degrees of freedom) of 47/163 
(compared to 71/165 without DM). Intriguingly the preferred mass of 
$m_{\rm DM} \simeq 60$\,GeV in the fit coincides with the narrow mass 
region around half of the Higgs mass which allows to reconcile existing 
constraints from direct detection, the relic density and collider searches, 
see~\cite{Cuoco:2017rxb} for a detailed analysis and discussion.

\section{Conclusion}\label{sec:summary} 

With the recently released AMS-02 data the cosmic-ray precision era has started. In order to 
reduce uncertainties with respect to the MIN/MED/MAX propagation scenarios we performed 
a joint fit of propagation and DM parameters. Intriguingly we found a feature in the antiproton 
flux which could be a possible hint for DM annihilation in the Galaxy.
We explore a wide class of annihilation channels including $gg$, $b \bar b$, $WW^{*}$, 
$ZZ^{*}$, $hh$ and $t\bar t$ which provide a statistical improvement of the fit at the 
4$\sigma$ level (3$\sigma$ level for $t \bar t$) with masses ranging from 35\,GeV to 
the Higgs- and top-production thresholds. A further investigation of these potential DM 
signals would have to examine systematic effects resulting from solar modulation, 
secondary antiproton production, and correlations in AMS-02 data.

Interestingly, for most channels the signal is largely compatible with a DM interpretation 
of the GCE as well as the constraints from dwarf spheroidal galaxies. 
As an example, we perform a joint fit for the specific case of the Higgs portal DM model
which points to a DM mass near half of the Higgs mass -- a region which is also favored 
by totally independent constraints, namely the interplay of the relic density and direct 
detection limits.

%

\end{document}